\documentclass[sigconf]{aamas} 

\usepackage{balance} 

\setcopyright{none}
\acmConference[ALA '22]{Proc.\@ of the Adaptive and Learning Agents Workshop (ALA 2022)}
{May 9-10, 2022}{Online, \url{https://ala2022.github.io/}}{Cruz, Hayes, Silva, Santos (eds.)}
\copyrightyear{2022}
\acmYear{2022}
\acmDOI{}
\acmPrice{}
\acmISBN{}
\acmSubmissionID{28002071}

\title[Hybrid teams in collective risk dilemmas]{The art of compensation: how hybrid teams solve collective risk dilemmas}

\author{Inês Terrucha}
\affiliation{
  \institution{IDLab, Ghent University – imec 
  \& Artificial Intelligence Lab, Vrije Universiteit Brussel
  }
  \city{Gent \& Brussels}
  \country{Belgium}
  }
\email{ines.terrucha@ugent.be}

\author{Elias Fernández Domingos}
\affiliation{
  \institution{ULB Machine Learning Group}
  \city{Brussels}
  \country{Belgium}}
\email{elias.fernandez.domingos@ulb.be }

\author{Francisco C. Santos}
\affiliation{
  \institution{INESC-ID \& Instituto Superior Técnico, Universidade de Lisboa}
  \city{Porto Salvo}
  \country{Portugal}}
\email{franciscocsantos@tecnico.ulisboa.pt}

\author{Pieter Simoens}
\affiliation{
  \institution{IDLab, Ghent University – imec}
  \city{Gent}
  \country{Belgium}}
\email{pieter.simoens@ugent.be}

\author{Tom Lenaerts}
\affiliation{
  \institution{ULB Machine Learning Group}
  \city{Brussels}
  \country{Belgium}}
\email{tom.leanerts@ulb.be}

\begin{abstract}
It is widely known how the human ability to cooperate has influenced the thriving of our species. However, as we move towards a hybrid human-machine future, it is still unclear how the introduction of AI agents in our social interactions will affect this cooperative capacity. Within the context of the one-shot collective risk dilemma, where enough members of a group must cooperate in order to avoid a collective disaster, we study the evolutionary dynamics of cooperation in a hybrid population made of both adaptive and fixed-behavior agents. Specifically, we show how the first learn to adapt their behavior to compensate for the behavior of the latter. The less the (artificially) fixed agents cooperate, the more the adaptive population is motivated to cooperate, and vice-versa, especially when the risk is higher. By pinpointing how adaptive agents avoid their share of costly cooperation if the fixed-behavior agents implement a cooperative policy, our work hints towards an unbalanced hybrid world. On one hand, this means that introducing cooperative AI agents within our society might unburden human efforts. Nevertheless, it is important to note that costless artificial cooperation might not be realistic, and more than deploying AI systems that carry the cooperative effort, we must focus on mechanisms that nudge shared cooperation among all members in the hybrid system.
\end{abstract}


\newcommand{\BibTeX}{\rm B\kern-.05em{\sc i\kern-.025em b}\kern-.08em\TeX}

\begin{document}


\pagestyle{fancy}
\fancyhead{}


\maketitle 

\section{Introduction}
\label{secIntroduction}


As Artificial Intelligence (AI) systems are making more and more decisions autonomously, we are relinquishing decision control, for example by allowing intelligent machines to accomplish some of our goals independently or alongside us (e.g., using Google translate to enable business opportunities across different languages \cite{hasyim2021human}), within the context of hybrid human-machine socio-technical systems (e.g., sharing the road with self-driving cars \cite{gopindra2021}). Given the extraordinary difficulties humans have demonstrated when trying to overcome global crises, such as the COVID-19 pandemic \cite{kunicova2020covid} or climate change \cite{milinski2008collective}, the question can be raised on how AI agents may help to resolve the problems in coordinating the efforts in those and similar mixed-motive situations. 


Even though many different works have advocated for the introduction of beneficial AI to promote human prosociality \cite{harvey2014hci,paiva2018engineering,oliveira2021towards}, others have pointed that humans may be keen to exploiting this benevolent AI behavior in their own favor \cite{camerer2018artificial,march2019behavioral,cohn2018honesty,algorithmexploitation}. Thus, before flooding society with AI applications with the promise that they could solve some of our most pressing issues, it is worth asking: What behavioral response can be expected in the presence of AI agent partners? How may decision-making potentially be affected? Will hybrid groups involving AI with predefined decision-processes actually achieve greater collective success? 

We frame here these questions within the context of the Collective Risk Dilemma (CRD) \cite{milinski2008collective}, a game that abstracts the conflict between helping the group to achieve a future goal at a personal cost, or free ride on the efforts of others and just collect the gains associated with achieving the goal. CRD is a public goods game with a delayed reward that is associated with societal problems like pandemic control through vaccination, climate negotiation to achieve CO$_{2}$ reduction and energy-grid usage by prosumers. Many experiments to assess human behavior have been performed \cite{milinski2008collective,tavoni2011inequality,milinski2011cooperative,dannenberg2015provision,cadsby1999voluntary}, where \cite{fernandez2021delegation} found that replacing all human participants by AI agents has a positive impact on the success rate of human groups.  Yet, they also showed that in hybrid groups of humans and AI agents this success is again reduced to the level of only human participants. Within this context, this manuscript aims to unravel in more detail the dynamics in hybrid human-agent groups, providing thus knowledge that allows one to design novel experiments to further this line of research on human-AI interactions in mixed-motive and competitive scenarios.

Using the two-action one-shot CRD as defined in~\cite{santos2011risk} (see also Methods), the current study aims to show which strategic responses can be expected (i.e., fraction of cooperative strategies) in groups consisting of AI agents with predefined stochastic behaviors and whether these learned responses are sufficient for success in reaching the goals. Hence two types of participants are considered, i.e. adaptive individuals that can change their behavior over time based on the outcome of their interactions (a proxy for potential human responses) and artificial agents that have a probabilistic behavior that does not change as a result of the interactions (a proxy for average AI agent behavior). We apply a social learning approach to alter the strategy of the first type of individuals. They can switch between the possible actions in function of their success in the interactions; both when there are other adapting individuals or stochastic AI agents in groups of a given size. Such social adaptation can be achieved in different ways (e.g. Roth-Erev learning \cite{macy2002learning}, Q-learning \cite{domingos2021modeling}), but here an evolutionary game theoretical approach is considered wherein strategic behaviors change in the adaptive individuals population by imitating those individuals that are performing the best \cite{hofbauer1998evolutionary,nowak2006evolutionary,traulsen2007pairwise,traulsen2009exploration,hindersin2019computation}.  

In our model, automation does not
assume that every AI agent will be continuously learning while
acting; rule-based systems are used in AI products and they
are hard-wired in the systems as learning on the fly might be
costly or even dangerous. Thinking about real-world AI applications, one should always consider that
producers of AI products want to give guarantees on what
the product does (also on what are its limitations, which is why we use a stochastic behavior that includes errors), and that allowing for extensive adaptation
while in use may be very risky. It is important to note that in this work, we are not considering the AI designer,
neither the dynamics involved behind them. We are simply
probing: If we consider this space of behaviors for AI agents,
what kind of human behaviors emerge given constant hybrid
interactions? Even though CRD scenarios are used to model very high risk events like a pandemic or the climate change, the same kind of non-linearity could be observed within any industrial or software hybrid team, which if the project is not delivered might suffer the consequences of losing their bonuses or even their jobs. This is especially interesting to probe since most teams are already hybrid if we consider the extensive productivity softwares that are already available in the market.

We will show that the adaptive individuals in this type of hybrid teams - where humans are informed about their artificial counterparts use and limitations - respond by exploiting the benevolence of the AI agents, by avoiding to contribute with cooperative efforts when the latter are already meeting the threshold needed, as previously hinted to in \cite{algorithmexploitation}. On the contrary, when the AI agents added to each group are associated with a lower capacity to contribute for the collective endeavor, and if the risk is high, the adaptive population cooperation levels are boosted to compensate. What is thus observed, is that in the presence of a set of pre-specified AI behaviors in a hybrid group of ``humans" that can adapt their behavior in function of their gains, one obtains compensating behavior in both directions, requiring thus additional mechanisms to align the efforts of both types of participants in the hybrid group dynamic. So, adding cooperative or pro-social agents into a group decision-making process with humans may increase the success of achieving the task, but it will not necessarily promote pro-social behavior.


\section{Related Work}
\label{secBackground}

 

In \cite{camerer2018artificial} it is pointed out that more experimental research is needed to really understand how human strategic decision-making changes when interacting with autonomous agents. Following on this, \cite{march2019behavioral} compiles a review of more than 90 experimental studies that have made use of computerized players. Its main conclusions validate that indeed, human behavior changes when some of the other players are artificial, and furthermore, the behavior deviates to become more rational (or in other words, selfish), where humans are observed to actually try to exploit the artificial players. 

This last conclusion was both supported by \cite{cohn2018honesty} and \cite{algorithmexploitation}. The first finds that humans cheat more against machines than against other humans, and thus prefer to play with machines, in an experiment that tested honesty in opposition to the possibility of higher financial gains. The latter recently published an experimental study that concludes that humans are keen on exploiting benevolent AI in various different classical social dilemma games. Within the context of the CRD used for the present work, \cite{fernandez2021delegation} groups participants in hybrid teams with AI agents. Even though 3 out of the 6 group members were AI agents that were successful in avoiding the risky outcome in previous treatments, the hybrid groups were not more successful than only human groups. Looking closer at the results, one can see that the average payoff of the humans in hybrid teams actually increases. These experimental results already hint towards the adoption of a compensatory behavior on part of the human members of the group once they are informed about the addition of somewhat collaborative agents to the group. 

In contrast with aforementioned works, \cite{harvey2014hci} and  \cite{paiva2018engineering} point towards the possibility of engineering prosociality in human behaviour through the use of pro-social AI agents. In the pursuit of this idea, \cite{oliveira2021towards} assembles a comprehensive review on the use of robots and virtual agents to trigger pro-social behaviour. Out of 23 studies included, 52\% reported positive effects in triggering such cooperative behavior. However, 22\% were inconclusive and 26\% reported mixed results. Moreover, while recent experimental works show that programming autonomous agents \cite{de2019human} that include emotion \cite{de2018shaping} or some form of communication \cite{crandall2018cooperating} may positively impact human cooperation, it is still unclear what are the mechanisms facilitating this effect. 

More directly related to our theoretical study, there are different works on the dynamics of how evolving populations adapt their behavioral profile given the introduction of agents with a fixed behavior (usually cooperative) either at the group level or at the population level \cite{vasconcelos2014climate,mao2017resilient,oakley2011pathological,santos2019evolution,shirado2017locally}. With our research questions, we also aim at understanding how the introduction of agents with a fixed behavior, not necessarily cooperative, affects the evolution of cooperation. 

\section{Methods}
\label{secMethodology}

\subsection{The one-shot Collective Risk Dilemma (CRD)}

In this manuscript we adopt the $N$ person one-shot CRD \cite{santos2011risk,santos2011risk,abou2012evolutionary,vasconcelos2013bottom,pacheco2014climate,hagel2016risk,domingos2020timing,domingos2021modeling}. Here, a group of $N$ individuals must each decide whether to Cooperate ($C$), by contributing a fraction $c$ of their initial endowment $b$, or to Defect ($D$) and contribute nothing. If the group contains at least $M$ $C$ players, i.e., the group contributes in total $Mcb$ ($M \leq N$) to the public good, then each player may keep whatever is left of their initial endowment. Otherwise, there is a probability $r$ that all players will loose all their savings and receive a payoff of $0$, hence the dilemma. Thus, the expected payoff of a $D$ and a $C$ player can be defined in functions of the number of $Cs$ in the group, $j$:

\begin{align}
    \pi_D(j) &= b(1-r + r\theta(j-M)) \label{piD} \\
    \pi_C(j) &= \pi_D - cb \label{piC},
\end{align}
\noindent where $\theta(x)$ is the Heaviside unit step function, with $\theta(x) = 0 $ if $x<0$ and $\theta(x) = 1 $ otherwise.  

\subsection{CRD with hybrid interactions}

We consider a population $H$ of $Z$ adaptive agents which are randomly sampled into groups of size $N-a$ to play the CRD with $a$ agents from population $A$ (whose individuals display a fixed averaged behavior).  This allows us, as explained in the section below, to investigate the population dynamics of this dilemma. 
When engaging in group interactions, each adaptive agent can either cooperate $C$ or defect $D$. The state of the population is then defined by the number of cooperators $k\in[0,Z]$. The behavior of the fixed agents is defined by their probability of cooperating in each interaction, $p\in[0,1]$, thus, they implement a stochastic (or mixed) strategy. 
In each group we can calculate the expected payoff of Ds or Cs in function of the number of cooperators from the adaptive population, $i$, the number of fixed agents $a$ and the payoff of a D (C) $\pi_{D(C)}$:

\begin{align}
    \Pi_{D(C)}(i,a,p) = p\pi_{D(C)}(j=i+a)+(1-p)\pi_{D(C)}(j=i)
\end{align}

  \begin{figure*}[ht!]
  \centering
  \includegraphics[width=0.85\textwidth]{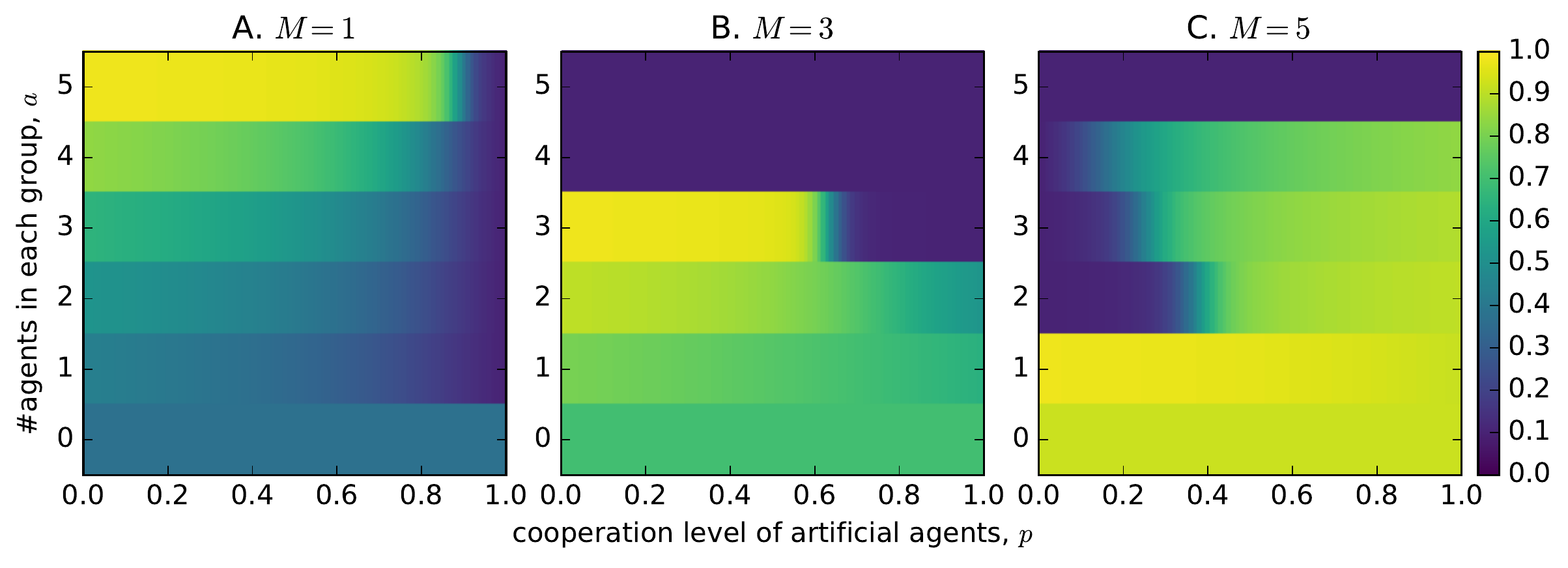}
  \caption{Average cooperation $\overline{C}$ (Eq.~\ref{EqavgCoop}) when interacting in groups of size $N=6$ with $a$ AI agents that cooperate with probability $p$. In order to avoid the risk of losing all the initial endowment with probability $r=0.9$ (value set for this figure), $M=1$ cooperators per group are needed in \textbf{A.}, $M=3$ in \textbf{B.} and $M=5$ in \textbf{C.}. When the threshold $M$ is low, the addition of enough highly cooperative agents discourages the need for increased more cooperation of the adaptive agents (see \textbf{A.} and \textbf{B.}). However, when the threshold $M$ is high, the addition of many highly cooperative AI agents turns the collective effort attainable and boosts cooperation (see \textbf{C.}). Other parameters used for this figure were: $Z=100$, $\mu=0.01$, $\beta=2$, $b=1$, $c=0.1$.}
  \label{figHumanCoopdiffMs}
\end{figure*}

The behavioral dynamics exhibited by the population of adaptive agents are governed by a social learning mechanism, where two randomly chosen individuals compare their fitness and imitate the one who is more successful within their social environment \cite{hofbauer1998evolutionary,nowak2006evolutionary,traulsen2007pairwise,traulsen2009exploration,hindersin2019computation}. Their fitness is the measure of the success of their current strategy (their payoff) averaged over all different group interactions. It can be defined as a function of the aforementioned variables by taking into account the population state and the payoffs given by Eqs.~\ref{piD} and \ref{piC}. Following on this, the fitness equations for cooperative (C) and defective (D) strategies, can be written as:
\begin{align}
    f_C &= \binom{Z-1}{N-a-1}^{-1} \sum_{i=0}^{N-a-1} \binom{k-1}{i}\binom{Z-k}{N-a-1-i}\Pi_{C}(i+1,a,p)  \label{fitC}
    \\
    f_D &= \binom{Z-1}{N-a-1}^{-1} \sum_{i=0}^{N-a-1} \binom{k}{i}\binom{Z-k-1}{N-a-1-i}\Pi_{D}(i,a,p).  \label{fitD}
\end{align}


Each agent in the population of adaptive agents may change its strategy profile at a given evolutionary step in the following way: an agent with a D (C) strategy is randomly selected from the population $H$ to adapt. With probability $\mu$ it will mutate into a C (D) strategy, otherwise, with probability $1-\mu$, it will compare its fitness with another randomly selected agent (assuming the newly selected agent has a different strategy)\cite{nowak2006evolutionary,hofbauer1998evolutionary,pacheco2009evolutionary,santos2011risk,traulsen2009exploration,traulsen2007pairwise,hindersin2019computation}. In case imitation is selected, a D (C) strategy will turn into a C (D) with a probability 
\begin{align}
    P(D\to C) = \frac{1}{1+e^{-\beta(f_C-f_D)}}, \label{fermi}
\end{align}
\noindent described by the Fermi function. This changes the state of the population $H$ of adaptive agents from $k$ to $k+1$. This probability becomes higher with a larger difference between the fitness of the two agents, $f_C-f_D$, or with a larger selection strength of the process, $\beta$. 

The transition probabilities that regulate the stochastic dynamics of population $H$, by defining the probability of increasing (+) or decreasing (-) the number of cooperators within a population are given by:
\begin{align}
    &T^+(k) = \frac{Z-k}{Z}\left((1-\mu)\frac{k}{Z-1}P(D\to C)\right) \label{Tplus}\\
    &T^-(k) = \frac{k}{Z}\left((1-\mu)\frac{Z-k}{Z-1}P(C\to D)\right), \label{Tminus}
\end{align}
\noindent where $P(C\to D)$ is obtained by replacing $C$ with $D$,  and $D$ with $C$ in Eq.~\ref{fermi}.

From these equations, we can construct the complete Markov chain of the $Z+1$ different states that fully describe the evolutionary process of the population $H$. From this Markov Chain we can compute the stationary distribution $P(k)$, the average cooperation level $\overline{C}$ and the average group success $\overline{s}_G$ of each population configuration. 

To compute the stationary distribution $P(k)$, we retrieve the eigenvector corresponding to the eigenvalue 1 of the tridiagonal transition matrix $S=[p_{ij}]^T$ \cite{nowak2006evolutionary,traulsen2009exploration,hindersin2019computation}. The values $p_{ij}$ are defined by the equations:
\begin{align}
    &p_{k,k\pm1} = T^{\pm}(k) \\
    &p_{k,k} = 1 - p_{k,k-1} - p_{k, k+1}
\end{align}
\noindent where the formulas that define $T^{\pm}(k)$ can be consulted in Eqs.~\ref{Tplus} and \ref{Tminus}.

From this it follows that the cooperation level $\overline{C}$ of population $H$ (for a given set of parameters $N$, $M$, $r$, $a$ and $p$) by averaging the fraction of cooperators in each population state, $k/Z$, over the stationary distribution of states $P(k)$ as given by:
\begin{align}
    \overline{C} = \sum_{k=0}^{Z}P(k)\frac{k}{Z}.
    \label{EqavgCoop}
\end{align}

As already mentioned, within the context of the CRD \cite{milinski2008collective,santos2011risk,abou2012evolutionary,vasconcelos2013bottom,pacheco2014climate,hagel2016risk,domingos2020timing,domingos2021modeling} another relevant quantity to derive is the probability of success of each group in reaching the threshold necessary of $M$ cooperators to avoid the collective risk.

At the population level, we compute the fraction of groups in each population state that are successful by resorting to the multivariate hypergeometric sampling, as follows
\begin{align}
    s_G(k) &= \binom{Z}{n-a}^{-1}\sum_{h=0}^{N-a}\binom{k}{h}\binom{Z-k}{N-h-a}\times \nonumber \\
     &\times \left(p\theta(h+a-M)+(1-p)\theta(h-M) \right) \label{successk}
\end{align}
\noindent where $\theta(x)$ is the Heaviside unit step-function as in Eqs.~1 and 2 of the main text.

Finally, similarly to what was done in Eq.~\ref{EqavgCoop}, we calculate the averaged group success by weighing the group success of each population state of Eq.~\ref{successk} over the stationary distribution of the evolutionary process $P(k)$:
\begin{align}
    \overline{s}_G = \sum_{k=0}^{Z}P(k)s_G(k).
    \label{EqavgSuccess}
\end{align}

\section{Results and Discussion}
\label{secResultsandDiscussion}

\subsection{Non-cooperative AI agents require the adapting population to cooperate to be successful}

What level of cooperation can be expected from the population of adapting individuals? Specifically, we want to know how average cooperation $\overline{C}$ (see Eq.~\ref{EqavgCoop}) varies in relation to four parameters: the number of AI agents ($a$) in each group, the probability of them cooperating ($p$), the probability of risking complete loss of initial endowment ($r$), and the number of cooperators needed per group to avoid that risk ($M$). \par

\begin{figure*}[ht!]
  \centering
  \includegraphics[width=0.65\textwidth]{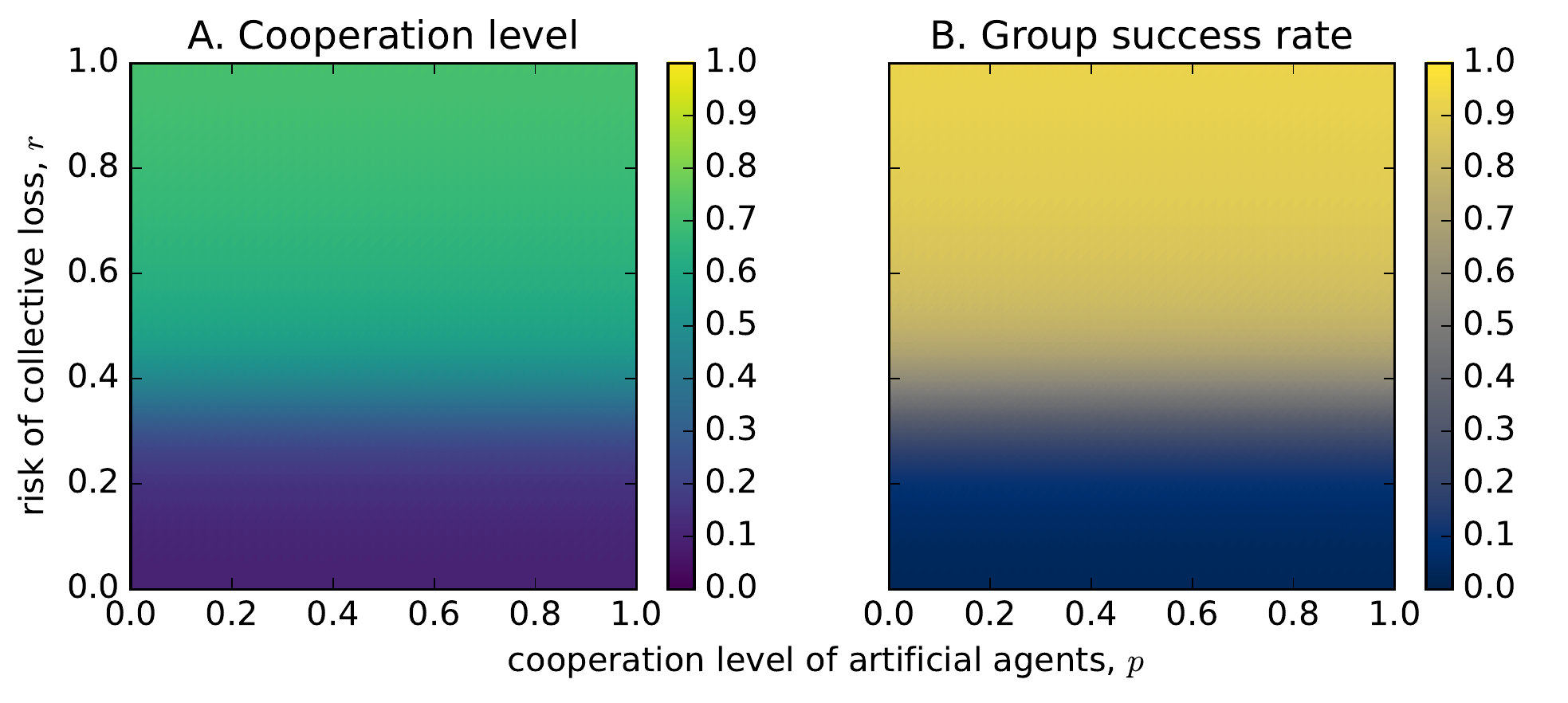}
  \caption{Control treatments with $a=0$ that accompany Figs.~\ref{figgroupSuccess} and ~\ref{fighumanCoop}. \textbf{A.} Average cooperation $\overline{C}$ (Eq.~9) of the population with adaptive individuals who engage in a collective risk dilemma only with members of the same population ($a=0$). \textbf{B.} Average group success $\overline{s}_S$ (Eq.~9) in avoiding the risk of collective loss for the population of adaptive individuals when interacting in groups without agents of fixed behavior ($a=0$). To produce these figures, we used: $N=6$, $M=3$, $Z=100$, $\mu=0.01$, $\beta=2$, $b=1$, $c=0.1$.}
  \label{figControl}
\end{figure*}

In Fig.~\ref{figHumanCoopdiffMs}, the variation in average cooperation $\overline{C}$ is shown in terms of the number of AI agents per group $a$ and their probability of cooperation $p$. Here, the group size is set to $N=6$, and the collective risk of loss to $r=0.9$ if the group is unable to reach a threshold of at least $M$ cooperators. $M=1$ (or $3$ or $5$) thus means that at least $1$ (or $3$ or $5$) participant needs to cooperate to avoid disaster. It is observed in Fig.~\ref{figHumanCoopdiffMs} that when the AI agents are at a fully cooperative behavior, i.e. $p=1$, there is no motivation for human cooperation once the number of agents is equal to the threshold value, $a=M$, as they already contribute with all the effort necessary to avoid collective risk of loss. On the contrary, when AI agents that contribute with probability $p=0$ to the avoidance of that risk are introduced, then the $N-a$ adaptive agents in the group are pressured to cooperate in order to reach $M$ by themselves. Still, if the threshold is high enough (in comparison to $N$), the introduction of at least somewhat cooperative agents makes the threshold $M$ more attainable, especially when we consider low costs of contribution as we do here with $c=0.1b$ (see Fig.~\ref{figHumanCoopdiffMs}C).\par 

\begin{figure*}[ht!]
  \centering
  \includegraphics[width=0.85\textwidth]{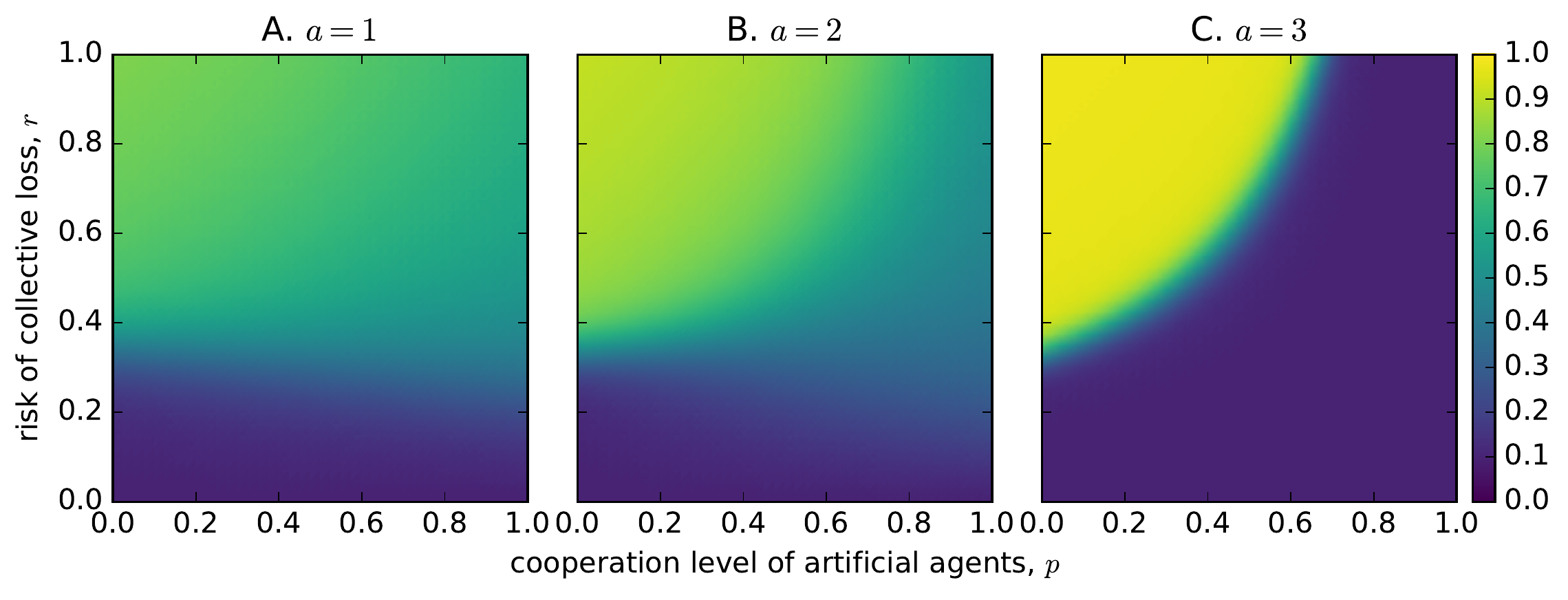}
  \caption{Average cooperation $\overline{C}$ (Eq.~\ref{EqavgCoop}) of an evolving population with $Z$ individuals who engage in a CRD where $a$ group members (AI agents) cooperate with probability $p$. In order to avoid the collective risk $r$ of loss, there must be at least $M$ cooperators, either adaptive $h$ and/or artificial $a$, in a group with size $N$. To obtain these results $N=6$ and $M=3$ were used. For \textbf{A.} $a=1$, for \textbf{B.} $a=2$ and for \textbf{C.} $a=M=3$. AI agents are added in each group. Cooperation is highest when risk is high, the number of AI agents not contributing in the group is high ($N-M = a$), so that $M$ can only be achieved through human cooperators $M=h$. Other parameters used for this figure were: $Z=100$, $\mu=0.01$, $\beta=2$, $b=1$, $c=0.1$.}
  \label{fighumanCoop}
\end{figure*}

Prior work on the CRD \cite{santos2011risk,santos2011risk,abou2012evolutionary,vasconcelos2013bottom,pacheco2014climate,hagel2016risk,domingos2020timing,domingos2021modeling}, here shown through Fig.~\ref{figControl} where we consider no presence of artificial agents, has interestingly shown that as the risk of disaster when not reaching the target increases, average cooperation also increases, while full cooperation (i.e. a population consisting of only $C$-strategists) is not achieved even for high risk situations (see panel A in Fig.~\ref{figControl}). In contrast, our findings portrayed in Fig.~\ref{fighumanCoop} shows that one can now observe that by adding exogenous AI agents (from a population $A$) of increasing cooperativeness (higher $p$), the average cooperation level of the adapting population decreases, especially for higher risk (i.e., $r$). Yet, for this region of high risk, when the added agents are non-cooperative, one can observe a higher level of cooperation from the adaptive population. \par 

This trend becomes clearly visible when increasing the number of AI agents in each interacting group, i.e. $a$, as can be observed when comparing panels A-C in Fig.~\ref{fighumanCoop}: The steepness of the boundary between full and almost null $\overline{C}$ is increased in relation to $p$. This transition in the cooperation level of the adapting population is maximized for $a=M$, as is clearly illustrated in Fig.~\ref{figHumanCoopdiffMs}. So when adding AI agents, the average investment in cooperation by the adapting population may decrease for both low risk and high-risk situations where the AI agents are mostly cooperative, but it has to be maximised when the AI agents are not that much into cooperation. \par
Overall, the findings in Fig.~\ref{figHumanCoopdiffMs} are consistent with  recent experimental work of \cite{algorithmexploitation} where algorithm exploitation is proven to be the main driver of the lack of cooperation with AI. Humans were found to act selfishly, by leaving the AI agents less well-off although not out of a competitive wish to end up better off than the machine. With our work we do indeed find that whenever the threshold is met completely by $a=M$ fully cooperative AI agents, the behavior selected for is less likely be cooperative. However, when the threshold is high enough, the introduction of benevolent AI agents selects for cooperative behavior so that  the risk of losing all the endowment is avoided.

\begin{figure*}[ht!]
  \centering
  \includegraphics[width=0.85\textwidth]{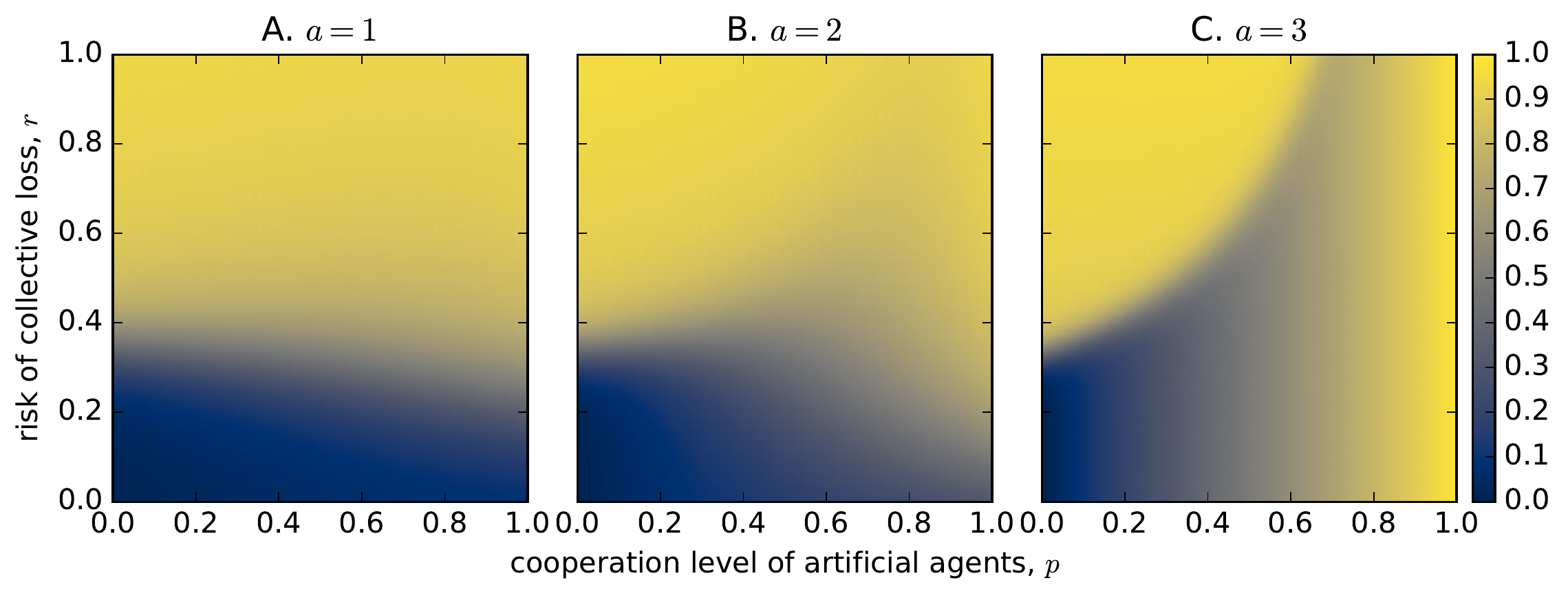}
  \caption{Average group success $\overline{s}_S$ (Eq.~\ref{EqavgSuccess}) in avoiding the risk of collective loss for each population configuration. In other words, probability of each group of size $N$ (here, $N=6$) having at least $M$ cooperators (here, $M=3$), averaged over every population state. In \textbf{A.} $a=1$, in \textbf{B.} $a=2$ and in \textbf{C.} $a=M=3$ AI agents are added in each group. \textbf{D.} $a=M=3$ artificial members are added in each group, which corresponds to the threshold of cooperators needed to avoid the collective risk. In \textbf{C.}, two boundaries can be clearly identified to achieve group success: the combination of a low AI agent cooperativeness and high risk of collective loss portray the upper left corner semi circle, and the presence of enough cooperative agents to achieve $M$ cooperators per group on their own draw a vertical line on the right of the graph. Overall, we can see that the introduction of AI agents increases the area of $r \times p$ where the threshold is met at least $90\%$ (painted in yellow) of the times. For this figure, $Z=100$, $\mu=0.01$, $\beta=2$, $b=1$, $c=0.1$.}
  \label{figgroupSuccess}
\end{figure*}

\subsection{Coordination and success in the CRD is achieved through compensation}

When studying a CRD, the average group success of that population in avoiding the collective risk, $\overline{s}_G$ (see Eq.~\ref{EqavgSuccess}), is even more essential, and of course deeply connected to the cooperation level observed in such a population. Fig.~\ref{figgroupSuccess} visualizes $\overline{s}_G$ for $N-a$ adaptive individuals and $a$ AI agents sampled from population $A$ that engage in a CRD with $N=6$, $M=3$ and varying $r$. Two regions are identified as responsible for high frequencies of group success: First, in Fig.~\ref{figgroupSuccess}A-C, we show that for high risk $r$ and low probability of cooperation ($p$) by the AI agents introduced in each group, the average success is high (with a frequency $>0.9$), which is consistent with the findings related to the cooperation level previously studied. Indeed, Fig.~\ref{figgroupSuccess} appears as result of the superposition between the cooperation level of the adaptive population shown in Fig.~\ref{fighumanCoop} and vertical lines that would represent the boundary where high cooperative effort from the fixed agents is sufficient to achieve group success.\par
Within the boundaries marked by low $p$, the boost in the average success rate is due to the evolved behavior in the population alone, since within this boundary the contribution of the AI agents to reach the threshold $M$ is minimum. Note also that the range for which cooperation is preferred is larger than what one would expect in a scenario where there are no AI agents (see Fig.~\ref{figControl}B). Second, for high values of $p$, the increase in average group success is especially justified by the introduction of fully cooperative AI agents ($p=1$). This effect is notably observable for values of $a$ closer to $M$, as one can see from Fig.~\ref{figgroupSuccess}C. With the increment of $a$, the average group success also becomes increasingly more dependent on $p$, especially in groups where $N-a < M$ and $a \geq M$, since the number of humans per group cannot change in any way the group outcome and the agents are already responsible for most of the possible cooperative effort.

\subsection{The effect of the number of AI agents a, the group size N and the threshold M}

The addition of AI agents to each group corresponds to a change in the game environment itself. Which is why the adaptive population is able to exploit such game transformation, benefiting from having information about the cooperation level of the added AI agents. Here we show how exactly the addition of AI agents with associated probability to cooperate ($p$) affects the stationary distribution of the population dynamics engaging in these hybrid group interactions.  

\begin{figure*}[t]
  \centering
  \includegraphics[width=0.85\textwidth]{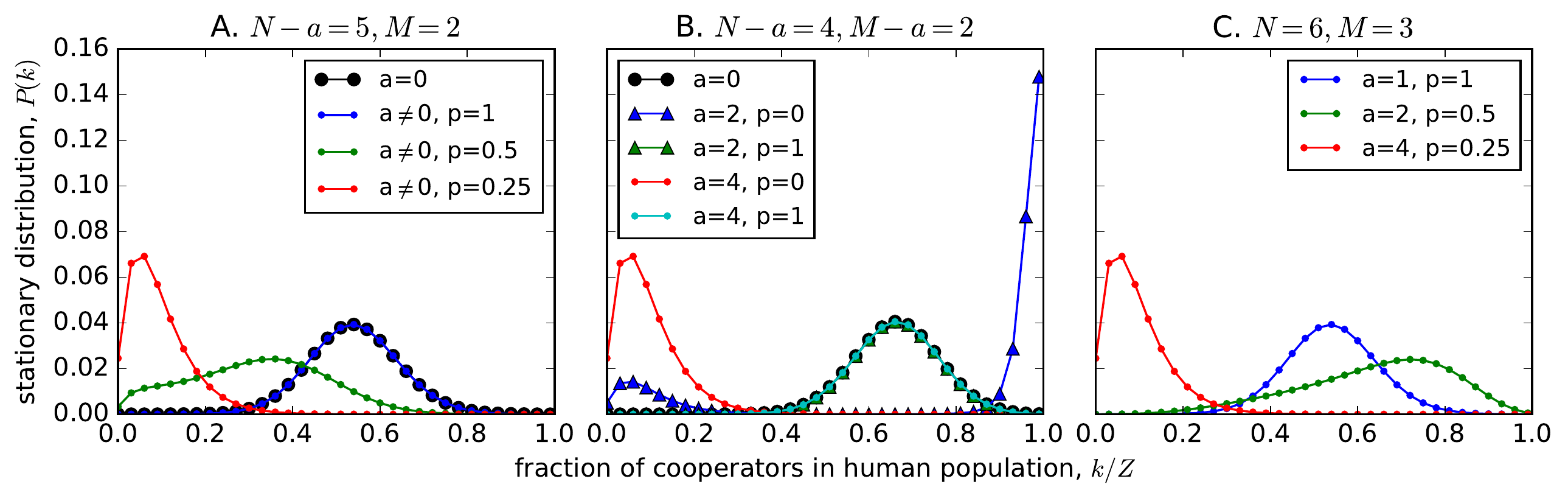}
  \caption{Stationary distribution in terms of fraction of evolving cooperators for different values of $N$, $M$, $a$ and $p$. In \textbf{A.} we set $N-a = 5$ and $M=2$, with results holding for any values of $N$ and $a$ that satisfy that condition. One can see that for fixed values of $N-a$ and $M$, the addition of AI agents with $p=0$ is equivalent to adding no AI agents, i.e. for $N-a = N$. In \textbf{B.} we fix $N-a =4$ and $M-a =2$. We find that fixing both $N-a$ and $M-a$ when $a\neq0$ and $p=1$ matches the results obtained for the case of $a=0$ with same $N-a$ and $M-a$. In \textbf{C.} we show how setting $a$ and $p$, so that $E = \sum pa$ is maintained, does not produce equivalent results. Specifically we plot the curves for $(a=1, p=1)$, $(a=2, p=0.5)$ and $(a=4, p=0.25)$, for groups of $N=6$ and $M=3$. Other parameters used were: $Z=100$, $\mu=0.01$, $\beta=2$, $b=1$, $c=0.1$ and $r=0.5$.}
  \label{figstatdist}
\end{figure*}

By introducing $a$ AI agents within a group, the group size $N$ is transformed into $N - a$, reducing thus the number of adapting individuals that can be introduced in each group, and - depending on whether the AI agents are cooperative with $p=1$ (or defective with $p=0$) - one also changes the number of those individuals needed to avoid the risk as $M  \to M - a$ (or $ M \to M $). This effect can be observed by looking at Figs.~\ref{figstatdist}A. and B. In Fig.~\ref{figstatdist}A, the values $N-a$ and $M$ were fixed to be able to observe these trends. Indeed, when adding $a$ AI agents that contribute nothing to reach threshold $M$ (with $p=0$) one can observe that it is the same as adding no AI agents, since $N-a > M$ and there is still the possibility for the cooperative members of the population to reach the threshold themselves. Furthermore, we find that for bigger values of $N-a$, cooperation becomes harder as the stationary distribution shifts to less cooperative states (similarly to the observed effect of increasing $N$ \cite{santos2011risk}). 

When the added AI agents are cooperative, the transformation $N \to N -a$ is also valid for changing the threshold $M \to M-a$. As it is shown in Fig.~\ref{figstatdist}B., the addition of a cooperative AI agents corresponds to both limiting the available places within the group and to lowering the threshold by that number of agents, who are contributing to the cooperative efforts themselves. 

However, when adding AI agents with a fixed non-deterministic behavior, i.e. for $0 < p < 1$, the resulting stationary distribution cannot be simply accounted for by a transformation in $N$ or in $M$, as is evinced for example in Fig.~\ref{figstatdist}A. Both the number of AI agents in each group and their cooperation level contribute to shape the resulting stationary distribution of the adaptive population. To support these conclusions one may look at Fig.~\ref{figstatdist}C., where we show the resulting stationary distributions for the same $N$, $M$ and $E=\sum pa$, where the latter should account for the cooperative effort employed by the addition of agents to the group. We find that even though the $E$ is fixed at $1$, different curves are obtained.  

Overall, Fig.~\ref{figstatdist} exacerbates the relevance of this study by pointing towards the complexity of adapting a population simply by fixing the behavior of some group members present in each interaction. Indeed, even though we are simply transforming the game environment, the resulting dynamics of the adaptive population are not trivial.

\section{Conclusions}
\label{secConclusions}

In this work we investigated what behavior is selected by social learning in the context of the one-shot CRD, when interactions occur in hybrid groups made of adapting agents (a proxy for human decision-making) and AI agents with fixed probabilistic behaviors (a proxy for average AI behavior). This model is used as a thought experiment to reason about the behavior one could expect in hybrid groups of humans and AI agents. It focuses on mixed-motive situations where there is a conflict between individual and common interests, as well as a risk that of not getting any benefits when not achieving the common goal, which affects the group as a whole. The model allowed us to understand under what conditions the introduction of AI agents in interacting groups actually contributes to an increase in their cooperation and their success rate in reaching a common goal. We have disentangled how the changes in success rate are related to the effort produced by the AI agents: whenever the latter are perceived as highly cooperative, the behavior of the adaptive population evolves to exploit the AI agents' benevolence. This effect was also shown by \cite{algorithmexploitation}. However, when the AI agents are low contributors, the adaptive population shifts to compensate those low contributions, as success can only be achieved when the goal is reached. Yet, this is only true for higher risk levels. In general, our model appears to indicate that the success rate of hybrid human-agents teams may be higher for a larger variety of settings than for human-only groups, or even for AI agents alone if they are not able to make the full contribution needed (lower $p$). Thus, our research suggests that there is potential benefit of using AI to nudge cooperation in human groups, which will need to be verified experimentally. Nevertheless, our findings also point towards an unbalanced future for human and AI teams. By working alongside cooperative AI, humans will eventually adapt to relax their own cooperative efforts. Hence, we must either identify AI policies that avoid this scenario and still promote cooperation to avoid collective risks or promote other modes of interaction in-between hybrid teams.



\begin{acks}
This research was funded by FWO-Vlaanderen under grant agreement no. G054919N. T.L benefits from the support by the Flemish Government through the AI Research Program and F.C.S and T.L. by TAILOR, a project funded by EU Horizon 2020 research and innovation program under GA No 952215. F.C.S. acknowledges support from FCT Portugal (UIDB/50021/2020, PTDC/MAT-APL/6804/2020 and PTDC/CCI-INF/7366/2020).
\end{acks}



\bibliographystyle{ACM-Reference-Format} 
\bibliography{biblio.bib}


\begin{thebibliography}{38}


\ifx \showCODEN    \undefined \def \showCODEN     #1{\unskip}     \fi
\ifx \showDOI      \undefined \def \showDOI       #1{#1}\fi
\ifx \showISBNx    \undefined \def \showISBNx     #1{\unskip}     \fi
\ifx \showISBNxiii \undefined \def \showISBNxiii  #1{\unskip}     \fi
\ifx \showISSN     \undefined \def \showISSN      #1{\unskip}     \fi
\ifx \showLCCN     \undefined \def \showLCCN      #1{\unskip}     \fi
\ifx \shownote     \undefined \def \shownote      #1{#1}          \fi
\ifx \showarticletitle \undefined \def \showarticletitle #1{#1}   \fi
\ifx \showURL      \undefined \def \showURL       {\relax}        \fi
\providecommand\bibfield[2]{#2}
\providecommand\bibinfo[2]{#2}
\providecommand\natexlab[1]{#1}
\providecommand\showeprint[2][]{arXiv:#2}

\bibitem[\protect\citeauthoryear{Abou~Chakra and Traulsen}{Abou~Chakra and
  Traulsen}{2012}]%
        {abou2012evolutionary}
\bibfield{author}{\bibinfo{person}{Maria Abou~Chakra} {and}
  \bibinfo{person}{Arne Traulsen}.} \bibinfo{year}{2012}\natexlab{}.
\newblock \showarticletitle{Evolutionary Dynamics of Strategic Behavior in a
  Collective-Risk Dilemma}.
\newblock \bibinfo{journal}{\emph{PLOS Computational Biology}}
  \bibinfo{volume}{8}, \bibinfo{number}{8} (\bibinfo{date}{08}
  \bibinfo{year}{2012}), \bibinfo{pages}{1--7}.
\newblock
\urldef\tempurl%
\url{https://doi.org/10.1371/journal.pcbi.1002652}
\showDOI{\tempurl}


\bibitem[\protect\citeauthoryear{Cadsby and Maynes}{Cadsby and Maynes}{1999}]%
        {cadsby1999voluntary}
\bibfield{author}{\bibinfo{person}{Charles~Bram Cadsby} {and}
  \bibinfo{person}{Elizabeth Maynes}.} \bibinfo{year}{1999}\natexlab{}.
\newblock \showarticletitle{Voluntary provision of threshold public goods with
  continuous contributions: experimental evidence}.
\newblock \bibinfo{journal}{\emph{Journal of Public Economics}}
  \bibinfo{volume}{71}, \bibinfo{number}{1} (\bibinfo{year}{1999}),
  \bibinfo{pages}{53--73}.
\newblock


\bibitem[\protect\citeauthoryear{Camerer}{Camerer}{2019}]%
        {camerer2018artificial}
\bibfield{author}{\bibinfo{person}{Colin~F. Camerer}.}
  \bibinfo{year}{2019}\natexlab{}.
\newblock \showarticletitle{Artificial Intelligence and Behavioral Economics}.
  In \bibinfo{booktitle}{\emph{The Economics of Artificial Intelligence: An
  Agenda}}. \bibinfo{publisher}{University of Chicago Press},
  \bibinfo{pages}{587--608}.
\newblock
\urldef\tempurl%
\url{http://www.nber.org/chapters/c14013}
\showURL{%
\tempurl}


\bibitem[\protect\citeauthoryear{Cohn, Gesche, and Mar{\'e}chal}{Cohn
  et~al\mbox{.}}{2018}]%
        {cohn2018honesty}
\bibfield{author}{\bibinfo{person}{Alain Cohn}, \bibinfo{person}{Tobias
  Gesche}, {and} \bibinfo{person}{Michel~Andr{\'e} Mar{\'e}chal}.}
  \bibinfo{year}{2018}\natexlab{}.
\newblock \bibinfo{booktitle}{\emph{Honesty in the digital age}}.
\newblock \bibinfo{type}{CESIfo Working Paper} 6996.
  \bibinfo{institution}{Center for Economic Studies and the ifo institute}.
  \bibinfo{pages}{1--26} pages.
\newblock
\showISSN{2364-1428}


\bibitem[\protect\citeauthoryear{Crandall, Oudah, Ishowo-Oloko, Abdallah,
  Bonnefon, Cebrian, Shariff, Goodrich, Rahwan, et~al\mbox{.}}{Crandall
  et~al\mbox{.}}{2018}]%
        {crandall2018cooperating}
\bibfield{author}{\bibinfo{person}{Jacob~W Crandall}, \bibinfo{person}{Mayada
  Oudah}, \bibinfo{person}{Fatimah Ishowo-Oloko}, \bibinfo{person}{Sherief
  Abdallah}, \bibinfo{person}{Jean-Fran{\c{c}}ois Bonnefon},
  \bibinfo{person}{Manuel Cebrian}, \bibinfo{person}{Azim Shariff},
  \bibinfo{person}{Michael~A Goodrich}, \bibinfo{person}{Iyad Rahwan},
  {et~al\mbox{.}}} \bibinfo{year}{2018}\natexlab{}.
\newblock \showarticletitle{Cooperating with machines}.
\newblock \bibinfo{journal}{\emph{Nature communications}} \bibinfo{volume}{9},
  \bibinfo{number}{1} (\bibinfo{year}{2018}), \bibinfo{pages}{1--12}.
\newblock


\bibitem[\protect\citeauthoryear{Dannenberg, L{\"o}schel, Paolacci, Reif, and
  Tavoni}{Dannenberg et~al\mbox{.}}{2015}]%
        {dannenberg2015provision}
\bibfield{author}{\bibinfo{person}{Astrid Dannenberg}, \bibinfo{person}{Andreas
  L{\"o}schel}, \bibinfo{person}{Gabriele Paolacci},
  \bibinfo{person}{Christiane Reif}, {and} \bibinfo{person}{Alessandro
  Tavoni}.} \bibinfo{year}{2015}\natexlab{}.
\newblock \showarticletitle{On the provision of public goods with probabilistic
  and ambiguous thresholds}.
\newblock \bibinfo{journal}{\emph{Environmental and Resource economics}}
  \bibinfo{volume}{61}, \bibinfo{number}{3} (\bibinfo{year}{2015}),
  \bibinfo{pages}{365--383}.
\newblock


\bibitem[\protect\citeauthoryear{de~Melo, Khooshabeh, Amir, and Gratch}{de~Melo
  et~al\mbox{.}}{2018}]%
        {de2018shaping}
\bibfield{author}{\bibinfo{person}{Celso~M de Melo}, \bibinfo{person}{Peter
  Khooshabeh}, \bibinfo{person}{Ori Amir}, {and} \bibinfo{person}{Jonathan
  Gratch}.} \bibinfo{year}{2018}\natexlab{}.
\newblock \showarticletitle{Shaping cooperation between humans and agents with
  emotion expressions and framing}. In \bibinfo{booktitle}{\emph{Proceedings of
  the 17th International Conference on Autonomous Agents and MultiAgent
  Systems}}. \bibinfo{pages}{2224--2226}.
\newblock


\bibitem[\protect\citeauthoryear{de~Melo, Marsella, and Gratch}{de~Melo
  et~al\mbox{.}}{2019}]%
        {de2019human}
\bibfield{author}{\bibinfo{person}{Celso~M de Melo}, \bibinfo{person}{Stacy
  Marsella}, {and} \bibinfo{person}{Jonathan Gratch}.}
  \bibinfo{year}{2019}\natexlab{}.
\newblock \showarticletitle{Human cooperation when acting through autonomous
  machines}.
\newblock \bibinfo{journal}{\emph{Proceedings of the National Academy of
  Sciences}} \bibinfo{volume}{116}, \bibinfo{number}{9} (\bibinfo{year}{2019}),
  \bibinfo{pages}{3482--3487}.
\newblock


\bibitem[\protect\citeauthoryear{Domingos, Gruji{\'c}, Burguillo, Kirchsteiger,
  Santos, and Lenaerts}{Domingos et~al\mbox{.}}{2020}]%
        {domingos2020timing}
\bibfield{author}{\bibinfo{person}{Elias~Fern{\'a}ndez Domingos},
  \bibinfo{person}{Jelena Gruji{\'c}}, \bibinfo{person}{Juan~C Burguillo},
  \bibinfo{person}{Georg Kirchsteiger}, \bibinfo{person}{Francisco~C Santos},
  {and} \bibinfo{person}{Tom Lenaerts}.} \bibinfo{year}{2020}\natexlab{}.
\newblock \showarticletitle{Timing uncertainty in collective risk dilemmas
  encourages group reciprocation and polarization}.
\newblock \bibinfo{journal}{\emph{iScience}} \bibinfo{volume}{23},
  \bibinfo{number}{12} (\bibinfo{year}{2020}), \bibinfo{pages}{101752}.
\newblock


\bibitem[\protect\citeauthoryear{Domingos, Gruji{\'c}, Burguillo, Santos, and
  Lenaerts}{Domingos et~al\mbox{.}}{2021a}]%
        {domingos2021modeling}
\bibfield{author}{\bibinfo{person}{Elias~Fern{\'a}ndez Domingos},
  \bibinfo{person}{Jelena Gruji{\'c}}, \bibinfo{person}{Juan~C Burguillo},
  \bibinfo{person}{Francisco~C Santos}, {and} \bibinfo{person}{Tom Lenaerts}.}
  \bibinfo{year}{2021}\natexlab{a}.
\newblock \showarticletitle{Modeling behavioral experiments on uncertainty and
  cooperation with population-based reinforcement learning}.
\newblock \bibinfo{journal}{\emph{Simulation Modelling Practice and Theory}}
  \bibinfo{volume}{109} (\bibinfo{year}{2021}), \bibinfo{pages}{102299}.
\newblock


\bibitem[\protect\citeauthoryear{Domingos, Terrucha, Suchon, Gruji{\'c},
  Burguillo, Santos, and Lenaerts}{Domingos et~al\mbox{.}}{2021b}]%
        {fernandez2021delegation}
\bibfield{author}{\bibinfo{person}{Elias~Fern{\'a}ndez Domingos},
  \bibinfo{person}{In{\^e}s Terrucha}, \bibinfo{person}{R{\'e}mi Suchon},
  \bibinfo{person}{Jelena Gruji{\'c}}, \bibinfo{person}{Juan~C Burguillo},
  \bibinfo{person}{Francisco~C Santos}, {and} \bibinfo{person}{Tom Lenaerts}.}
  \bibinfo{year}{2021}\natexlab{b}.
\newblock \showarticletitle{Delegation to autonomous agents promotes
  cooperation in collective-risk dilemmas}.
\newblock \bibinfo{journal}{\emph{arXiv e-prints}} (\bibinfo{year}{2021}),
  \bibinfo{pages}{arXiv--2103}.
\newblock


\bibitem[\protect\citeauthoryear{Hagel, Abou~Chakra, Bauer, and Traulsen}{Hagel
  et~al\mbox{.}}{2016}]%
        {hagel2016risk}
\bibfield{author}{\bibinfo{person}{Kristin Hagel}, \bibinfo{person}{Maria
  Abou~Chakra}, \bibinfo{person}{Benedikt Bauer}, {and} \bibinfo{person}{Arne
  Traulsen}.} \bibinfo{year}{2016}\natexlab{}.
\newblock \showarticletitle{Which risk scenarios can drive the emergence of
  costly cooperation?}
\newblock \bibinfo{journal}{\emph{Scientific reports}} \bibinfo{volume}{6},
  \bibinfo{number}{1} (\bibinfo{year}{2016}), \bibinfo{pages}{1--9}.
\newblock


\bibitem[\protect\citeauthoryear{Harvey, Golightly, and Smith}{Harvey
  et~al\mbox{.}}{2014}]%
        {harvey2014hci}
\bibfield{author}{\bibinfo{person}{John Harvey}, \bibinfo{person}{David
  Golightly}, {and} \bibinfo{person}{Andrew Smith}.}
  \bibinfo{year}{2014}\natexlab{}.
\newblock \showarticletitle{HCI as a means to prosociality in the economy}. In
  \bibinfo{booktitle}{\emph{Proceedings of the SIGCHI Conference on Human
  Factors in Computing Systems}}. \bibinfo{pages}{2955--2964}.
\newblock


\bibitem[\protect\citeauthoryear{Hasyim, Yolanda~Latjuba, Akhmar, Kaharuddin,
  and Jihad~Saleh}{Hasyim et~al\mbox{.}}{2021}]%
        {hasyim2021human}
\bibfield{author}{\bibinfo{person}{Muhammad Hasyim}, \bibinfo{person}{Ade
  Yolanda~Latjuba}, \bibinfo{person}{Andi~Muhammad Akhmar},
  \bibinfo{person}{Kaharuddin Kaharuddin}, {and} \bibinfo{person}{Noer
  Jihad~Saleh}.} \bibinfo{year}{2021}\natexlab{}.
\newblock \showarticletitle{Human-Robots And Google Translate: A Case Study Of
  Translation Accuracy In Translating French-Indonesian Culinary Texts}.
\newblock \bibinfo{journal}{\emph{Turkish Journal of Computer and Mathematics
  Education}} (\bibinfo{year}{2021}).
\newblock


\bibitem[\protect\citeauthoryear{Hindersin, Wu, Traulsen, and
  Garc{\'\i}a}{Hindersin et~al\mbox{.}}{2019}]%
        {hindersin2019computation}
\bibfield{author}{\bibinfo{person}{Laura Hindersin}, \bibinfo{person}{Bin Wu},
  \bibinfo{person}{Arne Traulsen}, {and} \bibinfo{person}{Julian Garc{\'\i}a}.}
  \bibinfo{year}{2019}\natexlab{}.
\newblock \showarticletitle{Computation and simulation of evolutionary game
  dynamics in finite populations}.
\newblock \bibinfo{journal}{\emph{Scientific reports}} \bibinfo{volume}{9},
  \bibinfo{number}{1} (\bibinfo{year}{2019}), \bibinfo{pages}{1--21}.
\newblock


\bibitem[\protect\citeauthoryear{Hofbauer, Sigmund, et~al\mbox{.}}{Hofbauer
  et~al\mbox{.}}{1998}]%
        {hofbauer1998evolutionary}
\bibfield{author}{\bibinfo{person}{Josef Hofbauer}, \bibinfo{person}{Karl
  Sigmund}, {et~al\mbox{.}}} \bibinfo{year}{1998}\natexlab{}.
\newblock \bibinfo{booktitle}{\emph{Evolutionary games and population
  dynamics}}.
\newblock \bibinfo{publisher}{Cambridge university press}.
\newblock


\bibitem[\protect\citeauthoryear{Karpus, Krüger, Verba, Bahrami, and
  Deroy}{Karpus et~al\mbox{.}}{2021}]%
        {algorithmexploitation}
\bibfield{author}{\bibinfo{person}{Jurgis Karpus}, \bibinfo{person}{Adrian
  Krüger}, \bibinfo{person}{Julia~Tovar Verba}, \bibinfo{person}{Bahador
  Bahrami}, {and} \bibinfo{person}{Ophelia Deroy}.}
  \bibinfo{year}{2021}\natexlab{}.
\newblock \showarticletitle{Algorithm exploitation: Humans are keen to exploit
  benevolent AI}.
\newblock \bibinfo{journal}{\emph{iScience}} \bibinfo{volume}{24},
  \bibinfo{number}{6} (\bibinfo{year}{2021}), \bibinfo{pages}{102679}.
\newblock
\showISSN{2589-0042}
\urldef\tempurl%
\url{https://doi.org/10.1016/j.isci.2021.102679}
\showDOI{\tempurl}


\bibitem[\protect\citeauthoryear{Kunicova}{Kunicova}{2020}]%
        {kunicova2020covid}
\bibfield{author}{\bibinfo{person}{Jana Kunicova}.}
  \bibinfo{year}{2020}\natexlab{}.
\newblock \showarticletitle{Driving the COVID-19 Response from the Center:
  Institutional Mechanisms to Ensure Whole-of-Government Coordination
  (English)}.
\newblock \bibinfo{journal}{\emph{Governance and Institutions Responses to
  COVID-19 Washington, D. C.}} (\bibinfo{year}{2020}).
\newblock


\bibitem[\protect\citeauthoryear{Macy and Flache}{Macy and Flache}{2002}]%
        {macy2002learning}
\bibfield{author}{\bibinfo{person}{Michael~W Macy} {and}
  \bibinfo{person}{Andreas Flache}.} \bibinfo{year}{2002}\natexlab{}.
\newblock \showarticletitle{Learning dynamics in social dilemmas}.
\newblock \bibinfo{journal}{\emph{Proceedings of the National Academy of
  Sciences}} \bibinfo{volume}{99}, \bibinfo{number}{suppl 3}
  (\bibinfo{year}{2002}), \bibinfo{pages}{7229--7236}.
\newblock


\bibitem[\protect\citeauthoryear{Mao, Dworkin, Suri, and Watts}{Mao
  et~al\mbox{.}}{2017}]%
        {mao2017resilient}
\bibfield{author}{\bibinfo{person}{Andrew Mao}, \bibinfo{person}{Lili Dworkin},
  \bibinfo{person}{Siddharth Suri}, {and} \bibinfo{person}{Duncan~J Watts}.}
  \bibinfo{year}{2017}\natexlab{}.
\newblock \showarticletitle{Resilient cooperators stabilize long-run
  cooperation in the finitely repeated prisoner’s dilemma}.
\newblock \bibinfo{journal}{\emph{Nature communications}} \bibinfo{volume}{8},
  \bibinfo{number}{1} (\bibinfo{year}{2017}), \bibinfo{pages}{1--10}.
\newblock


\bibitem[\protect\citeauthoryear{March}{March}{2019}]%
        {march2019behavioral}
\bibfield{author}{\bibinfo{person}{Christoph March}.}
  \bibinfo{year}{2019}\natexlab{}.
\newblock \bibinfo{booktitle}{\emph{{The Behavioral Economics of Artificial
  Intelligence: Lessons from Experiments with Computer Players}}}.
\newblock \bibinfo{type}{CESIfo Working Paper} 7926.
  \bibinfo{institution}{Center for Economic Studies and the ifo institute},
  \bibinfo{address}{Munich}. \bibinfo{pages}{1--40} pages.
\newblock
\showISSN{2364-1428}
\urldef\tempurl%
\url{https://ssrn.com/abstract=3485475}
\showURL{%
\tempurl}


\bibitem[\protect\citeauthoryear{Milinski, R{\"o}hl, and Marotzke}{Milinski
  et~al\mbox{.}}{2011}]%
        {milinski2011cooperative}
\bibfield{author}{\bibinfo{person}{Manfred Milinski}, \bibinfo{person}{Torsten
  R{\"o}hl}, {and} \bibinfo{person}{Jochem Marotzke}.}
  \bibinfo{year}{2011}\natexlab{}.
\newblock \showarticletitle{Cooperative interaction of rich and poor can be
  catalyzed by intermediate climate targets}.
\newblock \bibinfo{journal}{\emph{Climatic change}} \bibinfo{volume}{109},
  \bibinfo{number}{3} (\bibinfo{year}{2011}), \bibinfo{pages}{807--814}.
\newblock


\bibitem[\protect\citeauthoryear{Milinski, Sommerfeld, Krambeck, Reed, and
  Marotzke}{Milinski et~al\mbox{.}}{2008}]%
        {milinski2008collective}
\bibfield{author}{\bibinfo{person}{Manfred Milinski}, \bibinfo{person}{Ralf~D
  Sommerfeld}, \bibinfo{person}{Hans-J{\"u}rgen Krambeck},
  \bibinfo{person}{Floyd~A Reed}, {and} \bibinfo{person}{Jochem Marotzke}.}
  \bibinfo{year}{2008}\natexlab{}.
\newblock \showarticletitle{The collective-risk social dilemma and the
  prevention of simulated dangerous climate change}.
\newblock \bibinfo{journal}{\emph{Proceedings of the National Academy of
  Sciences}} \bibinfo{volume}{105}, \bibinfo{number}{7} (\bibinfo{year}{2008}),
  \bibinfo{pages}{2291--2294}.
\newblock


\bibitem[\protect\citeauthoryear{Nair and Bhat}{Nair and Bhat}{2021}]%
        {gopindra2021}
\bibfield{author}{\bibinfo{person}{Gopindra~S. Nair} {and}
  \bibinfo{person}{Chandra~R. Bhat}.} \bibinfo{year}{2021}\natexlab{}.
\newblock \showarticletitle{Sharing the road with autonomous vehicles:
  Perceived safety and regulatory preferences}.
\newblock \bibinfo{journal}{\emph{Transportation Research Part C: Emerging
  Technologies}}  \bibinfo{volume}{122} (\bibinfo{year}{2021}),
  \bibinfo{pages}{102885}.
\newblock
\showISSN{0968-090X}
\urldef\tempurl%
\url{https://doi.org/10.1016/j.trc.2020.102885}
\showDOI{\tempurl}


\bibitem[\protect\citeauthoryear{Nowak}{Nowak}{2006}]%
        {nowak2006evolutionary}
\bibfield{author}{\bibinfo{person}{Martin~A Nowak}.}
  \bibinfo{year}{2006}\natexlab{}.
\newblock \bibinfo{booktitle}{\emph{Evolutionary Dynamics: Exploring the
  Equations of Life}}.
\newblock \bibinfo{publisher}{Harvard University Press}.
\newblock


\bibitem[\protect\citeauthoryear{Oakley, Knafo, Madhavan, and Wilson}{Oakley
  et~al\mbox{.}}{2011}]%
        {oakley2011pathological}
\bibfield{author}{\bibinfo{person}{Barbara Oakley}, \bibinfo{person}{Ariel
  Knafo}, \bibinfo{person}{Guruprasad Madhavan}, {and}
  \bibinfo{person}{David~Sloan Wilson}.} \bibinfo{year}{2011}\natexlab{}.
\newblock \bibinfo{booktitle}{\emph{Pathological altruism}}.
\newblock \bibinfo{publisher}{Oxford University Press}.
\newblock


\bibitem[\protect\citeauthoryear{Oliveira, Arriaga, Santos, Mascarenhas, and
  Paiva}{Oliveira et~al\mbox{.}}{2021}]%
        {oliveira2021towards}
\bibfield{author}{\bibinfo{person}{Raquel Oliveira},
  \bibinfo{person}{Patr{\'\i}cia Arriaga}, \bibinfo{person}{Fernando~P Santos},
  \bibinfo{person}{Samuel Mascarenhas}, {and} \bibinfo{person}{Ana Paiva}.}
  \bibinfo{year}{2021}\natexlab{}.
\newblock \showarticletitle{Towards prosocial design: A scoping review of the
  use of robots and virtual agents to trigger prosocial behaviour}.
\newblock \bibinfo{journal}{\emph{Computers in Human Behavior}}
  \bibinfo{volume}{114} (\bibinfo{year}{2021}), \bibinfo{pages}{106547}.
\newblock


\bibitem[\protect\citeauthoryear{Pacheco, Santos, Souza, and Skyrms}{Pacheco
  et~al\mbox{.}}{2009}]%
        {pacheco2009evolutionary}
\bibfield{author}{\bibinfo{person}{Jorge~M Pacheco},
  \bibinfo{person}{Francisco~C Santos}, \bibinfo{person}{Max~O Souza}, {and}
  \bibinfo{person}{Brian Skyrms}.} \bibinfo{year}{2009}\natexlab{}.
\newblock \showarticletitle{Evolutionary dynamics of collective action in
  N-person stag hunt dilemmas}.
\newblock \bibinfo{journal}{\emph{Proceedings of the Royal Society B:
  Biological Sciences}} \bibinfo{volume}{276}, \bibinfo{number}{1655}
  (\bibinfo{year}{2009}), \bibinfo{pages}{315--321}.
\newblock


\bibitem[\protect\citeauthoryear{Pacheco, Vasconcelos, and Santos}{Pacheco
  et~al\mbox{.}}{2014}]%
        {pacheco2014climate}
\bibfield{author}{\bibinfo{person}{Jorge~M Pacheco},
  \bibinfo{person}{V{\'\i}tor~V Vasconcelos}, {and}
  \bibinfo{person}{Francisco~C Santos}.} \bibinfo{year}{2014}\natexlab{}.
\newblock \showarticletitle{Climate change governance, cooperation and
  self-organization}.
\newblock \bibinfo{journal}{\emph{Physics of life reviews}}
  \bibinfo{volume}{11}, \bibinfo{number}{4} (\bibinfo{year}{2014}),
  \bibinfo{pages}{573--586}.
\newblock


\bibitem[\protect\citeauthoryear{Paiva, Santos, and Santos}{Paiva
  et~al\mbox{.}}{2018}]%
        {paiva2018engineering}
\bibfield{author}{\bibinfo{person}{Ana Paiva}, \bibinfo{person}{Fernando~P
  Santos}, {and} \bibinfo{person}{Francisco~C Santos}.}
  \bibinfo{year}{2018}\natexlab{}.
\newblock \showarticletitle{Engineering pro-sociality with autonomous agents}.
  In \bibinfo{booktitle}{\emph{32nd AAAI Conference on Artificial
  Intelligence}}. \bibinfo{pages}{7994--7999}.
\newblock


\bibitem[\protect\citeauthoryear{Santos and Pacheco}{Santos and
  Pacheco}{2011}]%
        {santos2011risk}
\bibfield{author}{\bibinfo{person}{Francisco~C Santos} {and}
  \bibinfo{person}{Jorge~M Pacheco}.} \bibinfo{year}{2011}\natexlab{}.
\newblock \showarticletitle{Risk of collective failure provides an escape from
  the tragedy of the commons}.
\newblock \bibinfo{journal}{\emph{Proceedings of the National Academy of
  Sciences}} \bibinfo{volume}{108}, \bibinfo{number}{26}
  (\bibinfo{year}{2011}), \bibinfo{pages}{10421--10425}.
\newblock


\bibitem[\protect\citeauthoryear{Santos, Pacheco, Paiva, and Santos}{Santos
  et~al\mbox{.}}{2019}]%
        {santos2019evolution}
\bibfield{author}{\bibinfo{person}{Fernando~P Santos}, \bibinfo{person}{Jorge~M
  Pacheco}, \bibinfo{person}{Ana Paiva}, {and} \bibinfo{person}{Francisco~C
  Santos}.} \bibinfo{year}{2019}\natexlab{}.
\newblock \showarticletitle{Evolution of collective fairness in hybrid
  populations of humans and agents}. In \bibinfo{booktitle}{\emph{Proceedings
  of the AAAI Conference on Artificial Intelligence}},
  Vol.~\bibinfo{volume}{33}. \bibinfo{pages}{6146--6153}.
\newblock


\bibitem[\protect\citeauthoryear{Shirado and Christakis}{Shirado and
  Christakis}{2017}]%
        {shirado2017locally}
\bibfield{author}{\bibinfo{person}{Hirokazu Shirado} {and}
  \bibinfo{person}{Nicholas~A Christakis}.} \bibinfo{year}{2017}\natexlab{}.
\newblock \showarticletitle{Locally noisy autonomous agents improve global
  human coordination in network experiments}.
\newblock \bibinfo{journal}{\emph{Nature}} \bibinfo{volume}{545},
  \bibinfo{number}{7654} (\bibinfo{year}{2017}), \bibinfo{pages}{370--374}.
\newblock


\bibitem[\protect\citeauthoryear{Tavoni, Dannenberg, Kallis, and
  L{\"o}schel}{Tavoni et~al\mbox{.}}{2011}]%
        {tavoni2011inequality}
\bibfield{author}{\bibinfo{person}{Alessandro Tavoni}, \bibinfo{person}{Astrid
  Dannenberg}, \bibinfo{person}{Giorgos Kallis}, {and} \bibinfo{person}{Andreas
  L{\"o}schel}.} \bibinfo{year}{2011}\natexlab{}.
\newblock \showarticletitle{Inequality, communication, and the avoidance of
  disastrous climate change in a public goods game}.
\newblock \bibinfo{journal}{\emph{Proceedings of the National Academy of
  Sciences}} \bibinfo{volume}{108}, \bibinfo{number}{29}
  (\bibinfo{year}{2011}), \bibinfo{pages}{11825--11829}.
\newblock


\bibitem[\protect\citeauthoryear{Traulsen, Hauert, De~Silva, Nowak, and
  Sigmund}{Traulsen et~al\mbox{.}}{2009}]%
        {traulsen2009exploration}
\bibfield{author}{\bibinfo{person}{Arne Traulsen}, \bibinfo{person}{Christoph
  Hauert}, \bibinfo{person}{Hannelore De~Silva}, \bibinfo{person}{Martin~A
  Nowak}, {and} \bibinfo{person}{Karl Sigmund}.}
  \bibinfo{year}{2009}\natexlab{}.
\newblock \showarticletitle{Exploration dynamics in evolutionary games}.
\newblock \bibinfo{journal}{\emph{Proceedings of the National Academy of
  Sciences}} \bibinfo{volume}{106}, \bibinfo{number}{3} (\bibinfo{year}{2009}),
  \bibinfo{pages}{709--712}.
\newblock


\bibitem[\protect\citeauthoryear{Traulsen, Pacheco, and Nowak}{Traulsen
  et~al\mbox{.}}{2007}]%
        {traulsen2007pairwise}
\bibfield{author}{\bibinfo{person}{Arne Traulsen}, \bibinfo{person}{Jorge~M
  Pacheco}, {and} \bibinfo{person}{Martin~A Nowak}.}
  \bibinfo{year}{2007}\natexlab{}.
\newblock \showarticletitle{Pairwise comparison and selection temperature in
  evolutionary game dynamics}.
\newblock \bibinfo{journal}{\emph{Journal of theoretical biology}}
  \bibinfo{volume}{246}, \bibinfo{number}{3} (\bibinfo{year}{2007}),
  \bibinfo{pages}{522--529}.
\newblock


\bibitem[\protect\citeauthoryear{Vasconcelos, Santos, and Pacheco}{Vasconcelos
  et~al\mbox{.}}{2013}]%
        {vasconcelos2013bottom}
\bibfield{author}{\bibinfo{person}{Vitor~V Vasconcelos},
  \bibinfo{person}{Francisco~C Santos}, {and} \bibinfo{person}{Jorge~M
  Pacheco}.} \bibinfo{year}{2013}\natexlab{}.
\newblock \showarticletitle{A bottom-up institutional approach to cooperative
  governance of risky commons}.
\newblock \bibinfo{journal}{\emph{Nature Climate Change}} \bibinfo{volume}{3},
  \bibinfo{number}{9} (\bibinfo{year}{2013}), \bibinfo{pages}{797--801}.
\newblock


\bibitem[\protect\citeauthoryear{Vasconcelos, Santos, Pacheco, and
  Levin}{Vasconcelos et~al\mbox{.}}{2014}]%
        {vasconcelos2014climate}
\bibfield{author}{\bibinfo{person}{V{\'\i}tor~V Vasconcelos},
  \bibinfo{person}{Francisco~C Santos}, \bibinfo{person}{Jorge~M Pacheco},
  {and} \bibinfo{person}{Simon~A Levin}.} \bibinfo{year}{2014}\natexlab{}.
\newblock \showarticletitle{Climate policies under wealth inequality}.
\newblock \bibinfo{journal}{\emph{Proceedings of the National Academy of
  Sciences}} \bibinfo{volume}{111}, \bibinfo{number}{6} (\bibinfo{year}{2014}),
  \bibinfo{pages}{2212--2216}.
\newblock


\end{thebibliography}


\end{document}